\documentclass[11pt]{article}

\usepackage{graphicx}
\usepackage{amsmath}
\usepackage[myheadings]{fullpage}
\usepackage{natbib}
\usepackage{bbm}
\usepackage{url}
\usepackage[colorlinks = true,
            linkcolor = red,
            urlcolor  = blue,
            citecolor = blue,
            anchorcolor = blue]{hyperref}

\title{Detecting Multiple Random Changepoints in Bayesian Piecewise Growth Mixture Models}

\author{	ERIC F. LOCK\\
	\textit{Division of Biostatistics, School of Public Health,} \\
		\textit{University of Minnesota}\\[8pt]
	NIDHI KOHLI\\
	\textit{Department of Educational Psychology, College of Education \& Human
Development,}\\
	\textit{University of Minnesota}\\[8 pt]
		MAITREYEE BOSE\\
	\textit{Department of Biostatistics, School of Public Health,}\\
		\textit{University of Washington}\\
}
\date{}

\begin{document}
\maketitle
\begin{abstract}
Piecewise growth mixture models (PGMM) are a flexible and useful class of methods for analyzing segmented trends in individual growth trajectory over time, where the individuals come from a mixture of two or more latent classes.  These models allow each segment of the overall developmental process within each class to have a different functional form; examples include two linear phases of growth, or a quadratic phase followed by a linear phase.  The \emph{changepoint} (\emph{knot}) is the time of transition from one developmental phase (segment) to another.  Inferring the location of the changepoint(s) is often of practical interest, along with inference for other model parameters.   A random changepoint allows for individual differences in the transition time within each class.  The primary objectives of our study are: (1) to develop a PGMM using a Bayesian inference approach that allows the estimation of \emph{multiple} random changepoints within each class; (2) to develop a procedure to empirically \emph{detect the number of random changepoints} within each class; and (3) to empirically investigate the bias and precision of the estimation of the model parameters, including the random changepoints, via a simulation study.  We have developed the user-friendly package {\tt BayesianPGMM} for R to facilitate the adoption of this methodology in practice, which is available at \url{https://github.com/lockEF/BayesianPGMM}.  We describe an application to mouse-tracking data for a visual recognition task. 
\end{abstract}

\vspace{\fill}\pagebreak

\section{Introduction}

Longitudinal statistical models with piecewise functions (i.e., spliced lines or \emph{splines}) are extremely useful for analyzing data that demonstrate a segmented growth process, where the overall functional form incorporates two or more distinct functional forms with transitions from one phase to another over time \citep{kohli2015fitting}.  For instance, the development and/or decline in cognitive ability over time follows a piecewise trajectory (e.g., \citet{bradway1962intelligence,dominicus2008random,hall2001estimation,muniz2011random}).  The appeal of piecewise growth models is due to their flexibility to accommodate different kinds of functional forms with respect to each developmental phase (e.g., \citet{cudeck2002multiphase}).  To elaborate, a linear–-linear piecewise function can be used where the respective trajectory in the first and second developmental phase is linear, or a quadratic–-linear or exponential–-linear piecewise function can be used when the trajectory in the first developmental phase has some curvature and the rate of change in the second developmental phase is constant.  

An important parameter of a piecewise function is the \emph{changepoint} (\emph{knot}), the time of transition from one developmental phase (segment) to another.  The changepoint can be specified a priori or estimated as part of the model fitting procedure.  Inferring the location of the changepoint(s) is often of practical interest, along with inference for other model parameters.  It is possible to estimate a single changepoint or multiple changepoints under the assumption that transition times are common for a population of individuals (i.e., fixed changepoint), or that they vary from individual to individual (i.e., random changepoint) allowing for individual differences where the segments of the growth trajectories join (e.g., \citet{fearnhead2011efficient,lai2014identifying,morrell1995estimating,wang2008simulation}).
  
In many research settings, the data may consist of individuals from two or more unobserved subpopulations (i.e., classes).  For example, in studies related to the Head Start programs, it has been found that children with two or more years of program participation have slower achievement growth, on average, than children with only one year of program participation \citep{kreisman2003evaluating}.  Analysis of data with latent classes requires the extension of growth models to include a categorical latent grouping variable.  These statistical models are known as growth mixture models. The statistical framework of piecewise growth models can be extended to enable the identification of unknown classes, where individuals within a given class share similar mean segmented developmental trajectory with unknown changepoint(s) (e.g., \citet{kohli2013piecewise,kohli2015finite,kohli2015fitting}).  A piecewise growth mixture model (PGMM) can incorporate widely different changepoint locations and functional forms within each class; a PGMM with one class is simply referred to as a piecewise growth model (PGM).  For example, \citet{zhao2012bayesian} considered a PGMM with a single, unknown random changepoint to examine log prostate-specific antigen (PSA) monthly data after prostate cancer surgery, with race and cancer stage at diagnosis as covariates. 
 
The framework of PGMM is flexible and has a lot of utility in substantive research.  However, fitting these models is complicated and time intensive.  This is so because PGMMs are intrinsically nonlinear growth mixture models and the estimation of the mean or location of the changepoint(s) (a nonlinear parameter), along with the variance of the changepoint(s) is computationally very challenging.  Additionally, the parameter estimation for these models may either employ frequentist inference procedures (e.g., \citet{hall2000change,kohli2015finite,naumova2001tutorial}) or Bayesian inference procedures (e.g., \citet{carlin1992hierarchical,dominicus2008random}).  \citet{wang2008simulation}, and \citet{wang2009least} detailed the comparison of both inference approaches within the context of a single class piecewise growth model.  \citet{kohli2015fitting} compared the Bayesian-MCMC approach with the ML-EM approach for estimating PGMM with an unknown changepoint that is class specific but does \emph{not} vary for individuals within a class.  The results from their study showed Bayesian estimation generally performed better than frequentist ML-EM estimation, in terms of both the accuracy of the estimated model parameters and computational feasibility.  
 
 In all of these previous studies the researchers hypothesized and \emph{prefixed} the number of unknown changepoint locations, i.e., the number of changepoints were specified in advance.  There is no existing methodological study that empirically \emph{detects} the number of changepoints, i.e., considers the number of changepoints as unknown and to be inferred from the data, within a unified framework for inference.  This is limiting for many applications. Piecewise studies of educational data typically assume one changepoint \citep{sullivan2016longitudinal,kohli2015longitudinal,kieffer2012before}; however, it is plausible that many learning trajectories will have at least two changepoints: one preceding a period of accelerated growth (an ``a-ha" moment), and another preceding a period of decelerated growth (a ``saturation point") \citep{gallistel2004learning}.  Multiple changepoints are also plausible for many physical growth processes. It is generally agreed that human height growth occurs in at least three linear phases \citep{karlberg1987modelling}, but that these phases can be altered in malnourished subpopulations \citep{karlberg1994linear}.  For these and other applications a flexible inferential framework that allows for an arbitrary number of latent changepoints, as well as individual variation and population heterogeneity in the form of latent classes, is needed.

This leads to the primary objective of this article, that is, (1) to develop a PGMM using a Bayesian inference approach that allows the estimation of \emph{multiple} random changepoints within each class; (2) to develop a procedure to empirically \emph{detect} the number of random changepoints within each  class; and (3) to empirically investigate the bias and precision of the estimation of the model parameters, including the random changepoints, via a simulation study.  
        
The organization of this article is as follows.  In Section~\ref{model}, we introduce a general probabilistic model for a PGMM with random number of changepoints.  In Section~\ref{inference} we describe a generally applicable Bayesian approach to estimate the parameters of the model in Section~\ref{model}.  In Section~\ref{app} we describe an application to mouse-tracking data for a visual language processing experiment, using the Bayesian method described in Section~\ref{inference}.  In Section~\ref{sims} we describe  a comprehensive simulation study to assess the performance of the method, using results from the application in Section~\ref{app} as a starting point to simulate data.  In Section~\ref{impl} we describe an R package to apply the method. In Section~\ref{discussion} we discuss the advantages and limitations of our proposed approach, with potential extensions.  
        
\section{Likelihood model}       
\label{model}

Here we describe the data generating (likelihood) model for a piecewise growth mixture model with linear segments and latent number of changepoints.  Hierarchical prior distributions for unknown model parameters and other details that are specific to a Bayesian framework are discussed later in Section~\ref{inference}.  In what follows, Greek characters denote unknown parameters to be estimated.      

We first describe the  model for a single class and fixed number of changepoints.  For a single subject $i$, assume data are available for $M_i$ measurement occasions. For each occasion $j=1,\hdots,M_i$, let $y_{ij}$ denote the measured outcome for subject $i$ and let $x_{ij}$ denote the time of measurement.  The longitudinal trajectory of the outcome is characterized by $K+1$ segments of linear growth or decline, where $K$ is the number of changepoints. That is, for changepoint locations $\lambda_{i,1},\hdots,\lambda_{i,K}$,
\begin{align} \label{Eq1} y_{ij} = \beta_{i,0}+\beta_{i,1} x_{ij} + \sum_{k=1}^K \beta_{i,k+1} (x_{ij}-\lambda_{i,k})^+ + \epsilon_{ij},  \end{align}       
where $(\cdot)^+$ is the positive part function, \[(x_{ij}-\lambda_{i,k})^+ = \begin{cases} x_{ij}-\lambda_{i,k} &\mbox{ if } x_{ij}-\lambda_{i,k}>0 \\  0 &\mbox{ otherwise,}  \end{cases}\]  and the error terms $\epsilon_{ij}$ are independent and normally distributed with variance $\sigma_{\epsilon}^2$: \begin{align} \label{EqError} \epsilon_{ij} \overset{iid}{\sim} N(0,\sigma_{\epsilon}^2).\end{align} Because $\beta_{i,k+1}$ denotes the change in slope at the $k$'th changepoint, the slope of the $k$'th linear segment is given by $\beta_{i,1}+\beta_{i,2}+\hdots+\beta_{i,k}$. 

If subject $i$ belongs to a single-class cohort of size $N$, $i=1,\hdots,N$, it is natural to assume a Gaussian random effects model for the subject-specific intercepts, slope changes, and changepoints: 
\begin{align*}
\beta_{i,k} \overset{iid}{\sim} N(\bar{\beta}_k,\sigma_{\beta_k}^2) \, \, &\text{ for } k=0, ..., K+1, \text{ and} \\
\lambda_{i_,k}  \overset{iid}{\sim} N(\bar{\lambda}_k,\sigma_{\lambda_k}^2) \, \, &\text{ for } k=1, ... ,K, 	
\end{align*}
where $\{\bar{\beta}_k, \bar{\lambda}_k\}$ give the mean effects over all subjects and $\{\sigma_{\beta_k}^2,\sigma_{\lambda_k}^2\}$ give the variances of the subject-level effects.    For identifiability of the index $k$ we presume $\bar{\lambda}_1<\bar{\lambda}_2<...<\bar{\lambda}_K$.  

If the number of changepoints is unknown, we extend model (\ref{Eq1}) by introducing the latent parameter $\mathcal{K}$:
\begin{align}y_{ij} = \beta_{i,0}+\beta_{i,1} x_{ij} + \sum_{k=1}^K \beta_{i,k+1} (x_{ij}-\lambda_{i,k})^+ \mathbbm{1}_{\{k \leq \mathcal{K}\}}  + \epsilon_{ij},\label{eq3}\end{align}  
where $\mathbbm{1}_{\{\cdot\}}$ is the indicator 
\[\mathbbm{1}_{\{k \leq \mathcal{K}\}} = \begin{cases} 1 &\mbox{ if } k \leq \mathcal{K} \\  0 &\mbox{ otherwise}.  \end{cases}\] 
Note that $\mathcal{K} \in \{0,1,\hdots,K\}$ defines the number of changepoints, with a maximum of $K$ possible changepoints.

If the $N$ subjects belong to $C$ classes, we let $\psi(i) \in \{1,\hdots,C\}$ represent the class to which subject $i$ belongs for $i=1,\hdots,N$. In our context the class memberships $\psi(i)$ are unknown, and $\psi(i)=c$ with probability $\nu_c$ for $i=1,\hdots,N$ and $c=1,\hdots,C$.  Here $\nu_c$ defines the marginal probabilities in each class (i.e., the mixing proportions). The parameters $\{\beta_{i,k},\lambda_{i,k}\}$ arise from a $C$-component Gaussian mixture with class specific means and variances: 
\begin{align}
\label{Eq2}
\begin{split}
\beta_{i,k} \overset{iid}{\sim} N(\bar{\beta}_{\psi(i),k},\sigma_{\psi(i),\beta_k}^2) \, \, &\text{ for } k=0, ..., K+1, \text{ and} \\
\lambda_{i_,k}  \overset{iid}{\sim} N(\bar{\lambda}_{\psi(i),k},\sigma_{\psi(i),\lambda_k}^2) \, \, &\text{ for } k=1, ... ,K. 	
\end{split}
\end{align}  

The number of changepoints may also depend on the classes: 
\begin{align} \label{Eq3}  y_{ij} = \beta_{i,0}+\beta_{i,1} x_{ij} + \sum_{k=1}^K \beta_{i,k+1} (x_{ij}-\lambda_{i,k})^+ \mathbbm{1}_{\{k \leq \mathcal{K}_{\psi(i)}\}}  + \epsilon_{ij}, \end{align}
where $\mathcal{K}_1,\hdots,\mathcal{K}_C$ give the latent number of changepoints in each class.  The class membership  may either be known or latent.  

\section{Parameter inference}
\label{inference}
In this section we discuss inference for the unknown parameters of the model defined by Equations (\ref{EqError}), (\ref{Eq2}) and (\ref{Eq3}).  We use a Bayesian framework for inference.  Our reasons for adopting a Bayesian framework, rather than, for instance, a frequentist framework with maximum likelihood estimation, are several fold.  Bayesian inference gives a flexible philosophical framework for assessing uncertainty in all parameters, incorporating subjective prior beliefs, and borrowing information across multiple studies.  Furthermore, Bayesian inference has generally been shown to outperform maximum likelihood inference for less complex piecewise random effects models, in terms of both the accuracy of the estimated model parameters and computational feasibility \citep{wang2008simulation,kohli2015fitting}.  Moreover, maximum likelihood is generally prone to overfitting richly parametrized models \citep{myung2000importance}.  For example, a direct application of maximum likelihood to estimate the unknown parameters of the model given in Section~\ref{model} will always select the maximum number of changepoints ($\mathcal{K}_1=\hdots=\mathcal{K}_C=K$), because this case subsumes models where $\mathcal{K}_c < K$ for some class $c$. Alternatively, one could specify and fix $\{\mathcal{K}_c\}_{c=1}^C$ and estimate the remaining parameters via maximum likelihood.  Under this approach post-hoc penalized likelihood heuristics such as the Akaike information criterion (AIC) \citep{akaike1974new} or Bayesian information criterion (BIC) \citep{schwarz1978estimating} may be used to compare models with different candidate values for $\{\mathcal{K}_c\}_{c=1}^C$.  While AIC/BIC values can be used for model selection, they provide little intuition on the relative strength of evidence and uncertainty in the number of changepoints.  Moreover, AIC and BIC are justified asymptotically and their direct application is often inappropriate for models with multilevel or clustered effects (see, e.g., \citet{vaida2005conditional} and \citet{delattre2014note}), and estimating all permutations of the number of possible changepoints for multiple classes can be computationally intensive.  Rather, we treat $\{\mathcal{K}_c\}_{c=1}^C$ as random variables to be inferred in a unified Bayesian framework.

Let $D$ denote the set of all observed data 
\[D = \left\{(y_{ij},x_{ij}):i=1,\hdots,N \text{ and } j=1,\hdots,M_i \right\}\]
and let $\Theta$ denote the set of all unknown model parameters 
\begin{align}\label{params} \Theta = \left\{\sigma_\epsilon^2,\{\bar{\beta}_{c,k}\},\{\bar{\lambda}_{c,k}\},\{\sigma_{c,\beta_k}^2\},\{\sigma_{c,\lambda_k}^2\},\{\mathcal{K}_c\}, \nu \right\}.\end{align} 
The maximum number of possible changepoints $K$ and the maximum number of classes $C$ are fixed hyper-parameters.   Let $L(D \mid \Theta)$ be the multivariate Gaussian likelihood defined by Equations (\ref{EqError}), (\ref{Eq2}) and (\ref{Eq3}).  
 Under a Bayesian framework, one puts a \emph{prior} probability distribution on the parameter space, $p(\Theta)$.  An application of Bayes' rule gives the probability distribution of $\Theta$ conditional on the observed data:
\begin{align} \label{posterior} p(\Theta \mid D) = \frac{p(\Theta) L(D \mid \Theta)}{\int p(\Theta) L(D \mid \Theta) \, d\Theta} \,, \end{align}
this is called the \emph{posterior} probability distribution, and serves as the basis for parametric inference.  

\subsection{Prior selection}
\label{priors}

The choice of a prior distribution $p(\Theta)$ reflects the \emph{a priori} probability distribution of the parameters, which is refined by the observed data to give the posterior.  In practice the choice of a prior can be informed by previously observed data or expert opinion.  Alternatively, the prior can be non-informative, chosen to minimize subjective bias and maximize the influence of the observed data $D$ on the posterior.  We describe in detail a prior that is intended to be non-informative for a given maximum of changepoints ($K$) and number of possible classes ($C$).  This is the prior we use for our implementation, and it is designed to be applicable in a wide variety of situations by default when there is no clear rationale for a more informative prior.  Alternative priors may be used, depending on the application, especially when there is rationale to use a more subjective prior.  For example, here we construct a uniform discrete prior for the number of changepoints and a uniform prior over the inner range of $X$ for the location of the changepoints; if subject matter knowledge for a given application suggests more information on  the number and location of the changepoints,  this information can be expressed in an alternative prior.         

  We use the notation
\[X=\left\{x_{ij}:i=1,\hdots,N \text{ and } j=1,\hdots,M_i \right\}\] and 
\[Y=\left\{y_{ij}:i=1,\hdots,N \text{ and } j=1,\hdots,M_i \right\}.\]
Note that the measured time-points $x_{ij}$ need not be common across subjects, and the number of measurement occasions $M_i$ may depend on the subject $i$.  For example, if the measurement times $x_{ij}$ are generally common across subjects but some outcome measurements $y_{ij}$ are missing, time points with missing measurements for subject $i$ can be removed for subject $i$ only.   

The prior distribution for each model parameter is given with appropriate justification below.  For general applicability these priors should be robust to the scale of measurement for $X$ and $Y$.   Thus for some parameters the prior involves  \emph{empirical hyper-parameters}, i.e., parameters for a prior that depend on the data.   However, these hyper-parameters are given only in terms of the baseline mean and overall scale of the observed measurements, to ensure that they are invariant to the scale of measurement for a given application.  This is equivalent to using non-empirical priors after linearly scaling the data before analysis, in a way that will not affect the shape and structure of the trajectories (just scale the x- and y- axes).     

\begin{itemize}
\item $\sigma_\epsilon^2 \sim \mbox{Inverse-gamma}(0.001,0.001)$.  The inverse-gamma distribution is a \emph{conjugate} prior for the variance of a normal distribution, meaning both the prior and the posterior belong to the same distributional family (here, the inverse-gamma family).  The $\mbox{Inverse-gamma}(0.001,0.001)$ prior is a commonly used non-informative prior for the residual variance (see \citet{spiegelhalter1996}), as it accommodates both very small ($\sigma_\epsilon^2 \rightarrow 0$) and very large ($\sigma_\epsilon^2 \rightarrow \infty$) values.
         
\item $\bar{\beta}_{c,0} \overset{iid}{\sim} N(\bar{y}_{\cdot,1},\mbox{var}(y_{\cdot,1}))$ for $c=\{1,\cdots,C\}$, where $\bar{y}_{\cdot,1}$ and $\mbox{var}(y_{\cdot,1})$ are the sample mean and variance of outcomes at the initial time point.  The normal distribution is conjugate prior for the mean parameters of normally distributed random effects.  Here $\bar{y}_{\cdot,1}$ and $\mbox{var}(y_{\cdot,1})$ are  empirical hyper-parameters assuming that the initial time point corresponds to $0$ ($x_{i,1}=0$) for all subjects. If not, the time scale can be shifted so that the first measured time point corresponds to $x=0$.      
   
\item $\bar{\beta}_{c,k} \overset{iid}{\sim} N(0,(\mbox{sd}(Y)/\mbox{sd}(X))^2)$ for $c=\{1,\hdots,C\}$ and $k=\{1,\hdots,K+1\}$, where $\mbox{sd}(X)$ is the sample standard deviation of the time points and $\mbox{sd}(Y)$ is the sample standard deviation of the outcome over all time points. The empirical hyper-parameter $\mbox{sd}(Y)/\mbox{sd}(X)$ is motivated by the observation that for the simple model \begin{align} \label{simplemod} y_{ij}=\beta_{0i} + \beta_i x_{ij}+\epsilon_{ij} \, \text{ with } \,\beta_i \sim N(\bar{\beta},\sigma^2_\beta),\end{align} $|\bar{\beta}|< \mbox{sd}(Y)/\mbox{sd}(X)$;  thus, we generally expect slope parameters to have absolute value smaller than $\mbox{sd}(Y)/\mbox{sd}(X)$.
      
\item $\bar{\lambda}_{c,k} \overset{iid}{\sim} \mbox{Uniform}(a,b)$ for $c=\{1,\hdots,C\}$ and $k=\{1,\hdots,K\}$, where $a$ is the second-to-earliest measured time point in $X$ (over all subjects), and $b$ is the second-to-latest time point.  We choose not to let the mean changepoints have uniform support over the full range of $X$ ($\mbox{Uniform}(\mbox{min}(X),\mbox{max}(X)))$ to avoid model identifiability issues at the boundaries.  
 
\item $\sigma_{c,\beta_0} \overset{iid}{\sim} \mbox{Uniform}(0,\mbox{sd}(y_{\cdot,1}))$ for $c=\{1,\hdots,C\}$, where $\mbox{sd}(y_{\cdot,1})$ is the sample standard deviation of outcomes at the initial time point.  Here and elsewhere we use a uniform distribution over a range of plausible values for the standard deviation of latent subject-specific parameters (see \citet{gelman2006prior}).   The upper bound $\mbox{sd}(y_{\cdot,1})$ is motivated by the observation that the variance of the subject-specific intercepts will always be less than the variance of the observed values at time $0$.   

\item $\sigma_{c,\beta_k} \overset{iid}{\sim} \mbox{Uniform}(0,\mbox{sd}(Y)/\mbox{sd}(X))$ for $c=\{1,\hdots,C\}$ and $k=\{1,\hdots,K+1\}$.  The upper bound $\mbox{sd}(Y)/\mbox{sd}(X)$ is motivated by the observation that in model (\ref{simplemod}), $\sigma^2_{\beta} < \mbox{sd}(Y)/\mbox{sd}(X)$.

\item $\sigma_{c,\lambda_k} \overset{iid}{\sim} \mbox{Uniform}(0,b)$  where 
\[b = \frac{\mbox{max}(X)-\mbox{min}(X)}{4},\] for $c=\{1,\hdots,C\}$ and $k=\{1,2,\hdots,K\}$.  The upper bound $b$ is motivated by the observation that  anything larger than $b$ necessarily implies  plausible values of the random changepoint occur outside of the range of $X$.  That is, $\sigma_{c,\lambda_k}=b$ implies that  
\[P \left(\lambda_{i,k} \in \bar{\lambda}_{c,k}\pm \mbox{range}(X)/2 \right) \approx 0.95,\]
and so if the changepoint mean is at the midpoint, $\bar{\lambda}_{c,k}=(\mbox{max}(X)+\mbox{min}(X))/2$,  plausible values of the changepoint cover the entire range of $X$.  

\item $\mathcal{K}_c \overset{iid}{\sim} \mbox{Uniform}\{0,1,\hdots,K\}$ for $c \in \{1,\hdots,C\}$.  That is, we give equal prior probability to each possible number of changepoints in each class. In practice, this is accomplished by introducing auxiliary Bernoulli variables $I_{c,k}$ for $k=1,\hdots,K$, with 
    \[P(I_{c,k})=\frac{K-k+1}{K-k+2},\] then \[\mathcal{K}_c= \sum_{k=1}^K \prod_{k'=1}^k I_{c,k'}.\]
    As an alternative prior, we also consider $\mathcal{K}_c \overset{iid}{\sim}$Binomial$(K,p)$, where the binomial probability $p$ controls the expected number of changepoints and can be used to specify a more conservative prior.  The binomial prior is constructed with Bernoulli variables $I_{c,k}$ with $P(I_{c,k})=p$, where 
    \[\mathcal{K}_c= \sum_{k=1}^K I_{c,k}.\]        

\item $\psi(i)=c$ with probability $\nu_c$ for $i=1,\hdots,N$ and $c=1,\hdots,C$.  By default, we use a Dirichlet$(1,\hdots,1)$ prior for $(\nu_1,\hdots,\nu_C)$, which is uniformly distributed over the standard $C-1$ simplex ($\nu_1+\nu_2+\cdots+\nu_C=1$).  When $C=2$, the Dirichlet$(\alpha_1,\alpha_2)$ distribution is equivalent to a Beta$(\alpha_1,\alpha_2)$ distribution for $\nu_1$. Therefore the default prior corresponds to a Uniform$(0,1)$ prior on $\nu_1$, because Beta$(1,1)$ is Uniform$(0,1)$.  More generally, a Dirichlet$(\alpha_1,\hdots,\alpha_C)$ prior can be used.  When $\alpha_1 = \cdots = \alpha_C=\alpha^*$, smaller values of $\alpha^*$ suggest less parity in the class sizes;  this can lead to under-identified classes in the posterior, because the prior allows some classes to have negligible probabilities.  Larger values of $\alpha^*$ suggest more parity in the class sizes.      
\end{itemize}

\subsection{Posterior computation}
\label{postcomp}
The posterior distribution (\ref{posterior}) cannot be derived analytically because the integral in the denominator is not tractable.  Rather, we simulate draws from this posterior distribution via Markov chain Monte Carlo (MCMC).  In particular, we use the package {\tt rjags} for R \citep{rjags}, which performs Gibbs sampling in combination with other techniques to draw from the posterior distribution.  Gibbs sampling was first described in \citet{geman1984stochastic} and is widely used to simulate draws from a Bayesian posterior with several parameters.  We refer to \citet{casella1992explaining} for an accessible and thorough introduction to Gibbs sampling.  
Sampling results in a Markov chain $\{\Theta^{(t)}=(\theta_1^{(t)},\hdots,\theta_d^{(t)})\}_{t=1}^T$ containing $T$ dependent samples from the joint posterior distribution defined by $p(\Theta \mid D)$. 

The posterior sampling algorithm requires specifying initial parameter values $\Theta^{(0)}=\{\theta_1^{(0)},\hdots,\theta_d^{(0)}\}$.  The initial samples $\Theta^{(1)}, \Theta^{(2)},\hdots$ are dependent on $\Theta^{(0)}$, and so may not be representative of the posterior distribution.  Typically, samples before a certain number of iterations $T_0$ are ignored, and sometimes called the \emph{burn-in}.  Those samples after iterations $T_0$, $\{\Theta^{(T_0+1)},\hdots,\Theta^{(T)}\}$, are used for posterior inference.  We obtain a point estimate for each parameter $\theta_i$ by taking the mean of posterior samples 
\[\hat{\theta}_i = \frac{1}{T-T_0} \sum_{t=T_0+1}^T \theta_i^{(t)}.\] 
We obtain a $95\%$ \emph{credible interval} (CI) $[a_i,b_i]$ where $a_i$ is the $2.5$th percentile and $b_i$ is the $97.5$th percentile of $\{\theta_i^{(T_0+1)}, \hdots, \theta_i^{(T)}\}$; because the sample percentiles approximate quantiles of the posterior distribution, this gives
\[P(\theta_i \in [a_i,b_i] \mid D) \approx 0.95.\]
For discrete parameters, such as the number of changepoints $\mathcal{K}_c$ for each class, the relative frequency of each value over posterior samples approximates the posterior probability for that value.    

In practice we initialize the model in two stages. First we perform Gibbs sampling for a small number of iterations for the simplified model with no within-class variability, 
\[\sigma^2_{c,\beta_k} = 0 \text{ and } \sigma^2_{c,\lambda_k} = 0\]
for all $c,k$.  In this first stage initial values are generated from the prior (Section~\ref{priors}), and Gibbs sampling is much faster than for the full model with random effects.  The mean over iterations for parameters  $\left\{\sigma_\epsilon^2,\{\bar{\beta}_{c,k}\},\{\bar{\lambda}_{c,k}\}\right\}$, and the mode for parameters $\left\{\{\mathcal{K}_c\},\psi\right\}$, are then used as initial values for the full model. 
    
For accuracy of posterior inference it is important that simulated draws after the burn-in have negligible dependence on the initial values, and that enough samples are taken to adequately approximate the full posterior distribution.  If both of these conditions are satisfied, we say the algorithm has \emph{converged}.  We assess convergence by running multiple MCMC chains, from different initial values, in parallel.  If the algorithm has converged we expect results obtained from each chain, from samples after the burn-in, to be similar.  In particular we use the \emph{potential scale reduction factor} (PSRF) \citep{gelman1992inference}, which is the square root of the ratio of the total variation across chains over the average variation within chains, as a robust measure of convergence.  The multivariate PSRF \citep{brooks1998general} is a generalization of the PSRF for a single parameter, to assess convergence of a parameter vector.    A PSRF or multivariate PSRF less than $1.1$ or $1.2$ are common thresholds \citep{sinharay2004experiences}.  

When Gibbs sampling for a mixture model there is a risk of \emph{label switching}.  For example, what is labeled class $1$ in the beginning of the chain may be labeled class $2$ at the end of the chain, and vice versa.  To address label switching we apply the Equivalence Classes Representatives (ECR) algorithm \citep{papastamoulis2014handling} available from the {\tt label.switching} package for R \citep{papastamoulis2015label}, as a post-hoc step after posterior samples are collected. This algorithm separately permutes the class labels at each iteration to best match a specified \emph{pivot} labeling.  The pivot is initially given by the most common label for each sample across the iterations, and it is updated for repeated applications of the ECR algorithm until convergence.   In addition to label-switching of the classes, we must consider identifiability of the changepoint labels. For $\mathcal{K}_c^{(t)}>1$, we permute labels of the piecewise components so that $\bar{\lambda}_{c,i}^{(t)}<\bar{\lambda}_{c,j}^{(t)}$ for all $i<j<\mathcal{K}_c^{(t)}$, which is necessary for the components to be consistently labeled across the MCMC iterations. 

Note that correlations between the random effects $\{\{\beta_{i,k}\},\{\lambda_{i,k}\}\}$ are not explicitly modeled by the prior given in Section~\ref{priors}.  This choice was motivated by simplicity and practicality.   Conjugate priors for a correlation/covariance matrix that facilitate Gibbs sampling such as the inverse-Wishart prior are inflexible and biased, whereas more flexible priors inhibit computational speed and convergence (see, e.g., \citet{daniels1999nonconjugate}).  Alternatively, we compute the empirical sample correlation between each pair of random effects, averaged over each sampling iteration, as a post-hoc estimate.  For example, a point estimate for the correlation between the first changepoint and the initial slope for Class $1$ is given by

\[\frac{1}{T-T_0} \sum_{t=T_0+1}^{T} \mbox{cor}\{(\beta_{i,1}^{(t)},\lambda_{i,1}^{(t)}):\psi(i)=1\}.\]

\section{Application}
\label{app}
We apply the estimation approach detailed in Section~\ref{inference} to mouse-tracking data for a language recognition task, originally described in \citet{incera2016mouse}.  For this task, subjects were shown text of a given color, and instructed to move a computer mouse and click on the word corresponding to that color as quickly as possible. Thus the conditions were analogous to a Stroop task, with certain experimental modifications. For this illustration we focus on the control task, in which participants were shown meaningless text of the given color (e.g., ``XXXXXX" in red);  the study also included a congruent task in which participants were shown the word of the same color (e.g., ``red" in red) and an incongruent task in which participants were shown the word of a different color (e.g., ``blue" in red).  The x-coordinate of the mouse was tracked for each trial, on the range of $-100$ (away from the correct color) to $+100$ (toward the correct color).  The x-location was measured at $20$ms intervals for a period of $1000$ms (1 second).  The task was repeated for $16$ trials for each of $60$ subjects.  We consider the average x-location at each time point over the trials, yielding a sample size of $N=60$ with $M_i=50$ time points for each subject.

The position of the mouse is considered a proxy for the cognitive process of determining the word corresponding to the correct color.  Therefore, the data for each subject represents a longitudinal learning trajectory for a simple task on a very short time scale (1 second).  We are interested in modeling the pattern of this trajectory, and interpreting this pattern as it relates to the cognitive processes of recognition and action.  Intuitively, we expect these trajectories to have $2$ changepoints, that vary from subject to subject.  All trajectories begin at the origin, and are initially flat as the subjects process the image.  Thus the first changepoint has a critical interpretation as the ``decision point", when the subject recognizes the correct word and begins to move their mouse in that direction. For the second changepoint we hope to capture the ``conclusion point", after the subject has reached the correct word and settles on their decision. We estimate the two change-point model for one, two, or three possible classes.  The two-class model had the lowest deviance information criterion (DIC) \citep{spiegelhalter2002bayesian}, with a DIC of $15863$ (the DIC for the single class model was $15978$, and for the three class model was $15898$).  Thus in what follows we focus on the two-class results, which also had the clearest interpretation.     Of the subjects, $20$ were fluent only in English (monolingual), and $40$ were fluent in English and one other language (bilingual).    Thus, we also consider whether the identified classes are representative of monolingual and bilingual speakers.     

\begin{figure}[!h]
\begin{center}
\includegraphics[width = \textwidth]{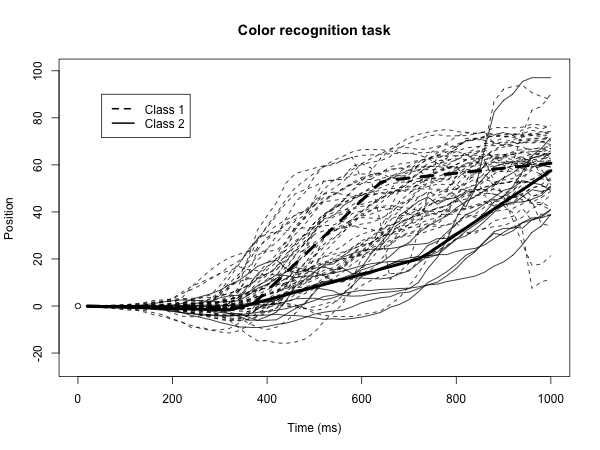}
\caption{Spaghetti plot of the mouse-tracking data for the color recognition task (a positive position corresponds to the correct color).  The trajectory defined by the mean parameters for each class are shown in \textbf{bold}. } 
\label{languagefig}
\end{center}
\end{figure}

\begin{figure}[!h]
\begin{center}
\includegraphics[width = \textwidth]{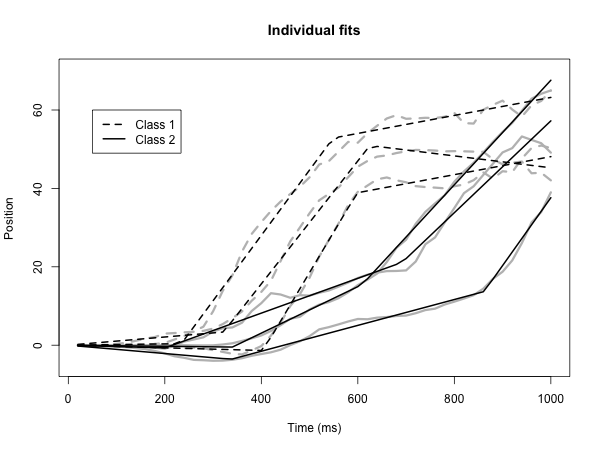}
\caption{Individual model fits for three representative subjects in Class 1 and three subjects in Class 2.  Raw data are shown in gray, piecewise model fits are shown in black.} 
\label{languagefigindiv}
\end{center}
\end{figure}

We run Gibbs sampling for the full model with $T=50,000$ total iterations, including $T_0=20,000$ iterations for burn-in, for $3$ MCMC chains. The sampling algorithm converges satisfactorily, with a PSRF across chains of less than $1.10$ for each of the parameters (\ref{params}), and a multivariate PSRF of $1.08$ over all parameters.  Table~\ref{tab1} gives parameter estimates and $95\%$ credible intervals for the overall and class-level parameters.  Figure~\ref{languagefig} shows a spaghetti plot showing the trajectory for each individual, and the class trajectories defined by the mean parameters for each class.  The class membership of each trajectory shown in Figure~\ref{languagefig} is determined by whichever class has the highest posterior probability for that individual.  As shown in Table~\ref{tab1}, there is considerable individual variability about the mean parameters for each class. Figure~\ref{languagefigindiv} shows the individual parameter fits for the model for three subject curves in each class;  these curves are chosen to be representative (with posterior probability $>0.975$ for their class) and to illustrate variation in the subject-level trajectories within each class.       

\begin{table}[!h]
\caption{Estimates, with 95\% credible interval, for model parameters in each class.}
\label{tab1}
\centering
\begin{tabular}{rrcrc}
  \hline
 Parameter & Class 1 estimate & Class 1 95\% CI & Class 2 estimate & Class 2 95\% CI\\ 
  \hline
$\nu$ & 0.77 & (0.53, 0.88) & 0.23 & (0.13, 0.47) \\ 
$\sigma_\epsilon$ & 3.16 & (3.08, 3.24) & 3.16 & (3.08, 3.24) \\ 
  $\beta_0$ & 0.003 & (-0.054, 0.054) & 0.000 & (-0.052, 0.053) \\ 
  $\beta_1$ & -0.002 & (-0.006, 0.001) & -0.005 & (-0.012, 0.002) \\ 
  $\beta_2$ & 0.194 & (0.172, 0.232) & 0.060 & (0.036, 0.108) \\ 
  $\beta_3$ & -0.171 & (-0.202, -0.143) & 0.081 & (-0.052, 0.135) \\ 
  $\sigma_{\beta_0}$ & 0.020 & (0.001, 0.038) & 0.019 & (0.001, 0.038) \\ 
  $\sigma_{\beta_1}$ & 0.010 & (0.008, 0.013) & 0.008 & (0.004, 0.014) \\ 
  $\sigma_{\beta_2}$ & 0.064 & (0.049, 0.082) & 0.027 & (0.013, 0.054) \\ 
  $\sigma_{\beta_3}$ & 0.079 & (0.056, 0.103) & 0.068 & (0.035, 0.122) \\ 
  $\lambda_{1}$ & 362 & (332, 391) & 321 & (222, 409) \\ 
  $\sigma_{\lambda_1}$ & 93.6 & (70.6, 119) & 132 & (72.2, 221) \\ 
  $\lambda_{2}$ & 643 & (567, 695) & 726 & (626, 827) \\ 
   $\sigma_{\lambda_2}$  & 149 & (102, 191) & 128 & (68.3, 222) \\ 
   \hline
\end{tabular}
\end{table}

The algorithm identifies one large class (Class 1; $\nu \approx 0.77$, or about $77\%$ of subjects) and one smaller class (Class 2).  There was strong evidence of two changepoints in each class, with posterior probability $P(\mathcal{K}_c=2 \mid D)$ above $0.99$ for $c=1,2$.  The mean trajectory of the larger class agrees with the intuition above, wherein the mouse curser is stationary at first, then increases sharply after the first changepoint when the subject makes a decision and moves in the correct direction, then is mostly stationary after the second changepoint after the subject settles on their decision.  However, for the smaller class there is a more gradual increase after the first changepoint, and a sharper increase after the second changepoint.  Our interpretation is that the subjects in Class 1 react quickly and confidently after making an initial decision, whereas subjects in Class 2 proceed more tentatively at first and gradually move toward their decision with increasing confidence.  Interestingly, the decision point, given by the first changepoint, is similar for both classes; it is the swiftness with which they react after their initial decision, and the pattern of their reaction, that distinguishes the two classes.

The $95\%$ credible intervals in Table~\ref{tab1} are relatively wide for the class proportion $\nu_2$, and for certain parameters in Class 2.  This is partly due to uncertainty in the class memberships: $8$ individuals had posterior class probabilities between $0.1$  and $0.9$. Uncertainty in the class memberships can drive uncertainty in the parameter estimates, particularly for the smaller class (Class 2).  Moreover, even for given class memberships, the flat uniform prior will in general yield more posterior variance for the underlying class proportion than, e.g., a more informative Dirichlet prior with larger hyperparameters.  

Of the $11$ subjects that were allocated to the larger class, $5$ were monolingual speakers.  The association between the identified class for each subject and monolingual/bilingual status was not significant (p-value$=0.48$, Fisher's exact test), suggesting that monolingual/bilingual status was not the dominant distinguishing characteristic in these data.       

\section{Simulation}
\label{sims}

Here we describe a comprehensive simulation study to evaluate different aspects of the proposed method, using the results from the application in Section~\ref{app} as a starting point.  We simulate data under the piecewise linear model of Section~\ref{model} to assess recovery of the true parameters, using the estimates in Table~\ref{tab1} as parameter values, with certain manipulated conditions.  We consider different values of the sample size $N$, and number of time points $M_i$, to assess how either affect the overall accuracy of the fitted model.  We consider different class proportions $\nu_c$, to assess how the relative sizes of each cluster affect class allocation.  We also manipulate the number of changepoints $\mathcal{K}_c$, as detecting the number of changepoints is a key innovation of the present model.   Specifically, we simulate under the following assumptions:  
\begin{itemize}
\item The overall sample size, $N$, is $60$ or $120$.
\item The number of time-points, $M_i$, is $25$ or $50$. For each simulation scenario the number of measurement occasions $M_i$ is the same for each subject $i=1,\hdots,N$.  If $M_i=25$, the measurement occasions occur after intervals of length $40$, $x_i=(40,80, 120, \hdots,1000)$, rather than intervals of length $20$ for $M_i=50$, $x_i=(20,40,60,\hdots,1000)$.   
\item The proportion of samples in Class 1, $\nu_1$, is $0.8$ or $0.5$.  If $\nu_1=0.8$ the classes are of size $48$ and $12$ for $N=60$ or $96$ and $24$ for $N=120$, if $\nu_1=0.5$ the classes are of size $30$ and $30$ for $N=60$ or $60$ and $60$ for $N=120$ .
\item The number of changepoints in Class 2, $\mathcal{K}_2$, is $0$, $1$, or $2$.  For $\mathcal{K}_2=1$ the effect at changepoint $2$ is removed, and for $\mathcal{K}_2=0$ the effects at both changepoints are removed, but other parameters for Class $2$ remain the same.  We include both changepoints for Class $1$ ($\mathcal{K}_1=2$) for all scenarios.    	
\end{itemize}   
We implement a fully crossed factorial  design with the above conditions, resulting in $2 \times 2 \times 2 \times 3 = 24$ simulation scenarios.  We generate $100$ replicated datasets under each simulation scenario, and  apply the estimation algorithm outlined in Section~\ref{inference} to each replication.

For each of the $2,400$ total replications, we ran the algorithm for $50,000$ sampling iterations per chain for each of the $2,400$ total replications, with a $20,000$ iteration burn-in; we ran three chains for each replication, to assess convergence.  The percent of replications that converged, with a mean PSRF over all parameters less than $1.2$, ranged from $26\%$ to $82\%$ across the $24$ simulation scenarios.  To assess the effect of the number of sampling iterations on convergence we repeated all $100$ replications in the application-motivated scenario with $N=60$, $M_i=50$, $\nu_1=0.80$, and $\mathcal{K}_2=2$ for a larger number of iterations.  For this scenario $44\%$ of replications converged within the first $50,000$ iterations (with $20,000$ burn-in), $76\%$ converged when the algorithm was run for $50,000$ more iterations ($100,000$ total; $50,000$ burn-in), and $91\%$ converged when the algorithm was run for $100,000$ more iterations ($200,000$ total; $100,000$ burn-in). These results suggest that poor convergence can be overcome by increasing the number of sampling iterations for certain situations.  

Among converged replications across all simulation conditions, the classes were generally well identified; when class membership was determined based on whichever class had higher posterior probability,  the mean misclassification rate across all replications was $3.9\%$.  The number of changepoints in each class was also generally well-recovered;  the average estimated posterior probability for the true number of changepoints, across all classes and simulation conditions, was $0.958$.   

We used multi-factor ANOVA to investigate the effects of the manipulated simulation conditions on three key performance metrics representing convergence, misclassification, and recovery of the number of changepoints.  For each metric we performed ANOVA with main effects for each of the four manipulated conditions.  Table~\ref{tab3} gives the significance of each effect, as well as the overall mean within each level.  Interestingly, while results generally improved with a larger sample size ($N=120$ vs. $N=60$), the number of measurement occasions ($M_i=50$ vs. $M_i=25$) did not significantly affect any metric.  

\begin{table}[ht]
\caption{ANOVA p-values and mean effects for four different simulations conditions (sample size $N$, measurement occasions $M_i$, class proportion $\nu_1$, and number of changepoints in Class 2 $\mathcal{K}_2$) on three different performance metrics.}
\label{tab3}
\centering
\begin{tabular}{|r|rrr|}
\hline
& Log$_2$(PSRF) & Misclassification & P(Correct $\mathcal{K}_c$)\\	
\hline
$N=60$ & 0.290 & 0.043& 0.954 \\
$N=120$ & 0.366 & 0.035& 0.962\\
\textbf{p-value}       & $\mathbf{<0.001}$ & $\mathbf{0.001}$ &  $\mathbf{0.256}$ \\
\hline 
$M_i=25$ & 0.334 & 0.041 & 0.957 \\
$M_i=50$ & 0.321& 0.038 & 0.958 \\
\textbf{p-value}        & $\mathbf{0.880}$ &  $\mathbf{0.307}$ &  $\mathbf{0.896}$\\
\hline 
$\nu_1=0.8$ & 0.320 & 0.056 & 0.917 \\
$\nu_1=0.5$ &0.336 & 0.025 &0.993\\
 \textbf{p-value}       &  $\mathbf{0.001}$ &  $\mathbf{<0.001}$ &  $\mathbf{<0.001}$ \\
\hline 
$\mathcal{K}_2=2$ & 0.464  & 0.071 & 0.985  \\
$\mathcal{K}_2=1$ & 0.334 & 0.061 & 0.902 \\
$\mathcal{K}_2=0$ & 0.185 & 0.005 & 0.979 \\
\textbf{p-value} & $\mathbf{<0.001}$ &  $\mathbf{<0.001}$&  $\mathbf{<0.001}$\\
\hline 
\end{tabular}
\end{table}

For Class $1$, which always had $2$ changepoints, the correct number of changepoints had posterior probability greater than $0.5$ for all simulation replications, and had posterior probability greater than $0.95$ for $99.8\%$ of replications.  For Class $2$, the mean probability for each number of changepoints is shown against the true number of changepoints in Table~\ref{tab4}. Here, too, the number of changepoints is generally well recovered.  However, the scenario with $1$ true changepoint gave non-trivial probability to the $2$ changepoint model;  this is likely in part due to misclassification, in which some of the subjects with $2$ changepoints (from Class $1$) are misallocated to Class $2$.     

\begin{table}[ht]
\caption{True number of changepoints and mean posterior probability, for Class $2$.}
\label{tab4}
\centering
\begin{tabular}{|r|rrr|}
\hline
& True $\mathcal{K}_2=2$ & True $\mathcal{K}_2=1$ & True $\mathcal{K}_2=0$\\	
\hline
Posterior $\mathcal{K}_2=2$ & 0.97 & 0.17& 0.00 \\
Posterior $\mathcal{K}_2=1$ &0.02 & 0.81& 0.04\\
Posterior $\mathcal{K}_2=0$ & 0.01 & 0.02 & 0.96\\
\hline
\end{tabular}
\end{table}

   Table~\ref{tab2} shows the bias and variability, as well as coverage rates for the 95\% credible interval, for the converged replications across all simulation conditions.  The parameters are generally well-recovered, but some are estimated with bias.  This bias is apparently due in part to misclassification; parameter estimates tend to shrink toward the midpoint of both classes, as a small proportion of subjects in Class 1 are misallocated to Class 2 and vice-versa.  Note also that the estimates for Class $2$ are generally less precise, due to its smaller sample size in several simulation scenarios. For the $N=601$ converged replications when $\nu_1 = 0.5$, the estimates for $\nu_1$ had mean $0.492$ and standard deviation $0.049$, yielding a z-standardized bias of $-0.16$.  For the $N=525$ converged replications when $\nu_1 = 0.8$, the estimates for $\nu_1$ had mean $0.77$ and standard deviation $0.079$, yielding a z-standardized bias of $0.295$.  
   
\begin{table}[!h]
\caption{Summary of parameter estimates in Class 1 and 2 over all converged replications ($1126$ total replications), including true values (in \textbf{bold}), mean of the estimates, standard deviation of the estimates across simulation replications, standardized bias (Z-bias=(Mean-Truth)/SD) and coverage rate of the $95\%$ credible interval.}
\label{tab2}
\centering
\begin{tabular}{rrrrrrrrrrr}
  \hline
& \textbf{Class 1} & Mean & SD & Z-bias & Coverage & \textbf{Class 2} & Mean & Sd & Z-bias & Coverage \\ 
  \hline
$\sigma_\epsilon$ & \textbf{3.16} & 3.17 & 0.054 & 0.28 & 0.929 & \textbf{3.16} & 3.17 & 0.054 & 0.28 & 0.929 \\ 
  $\beta_1$ & \textbf{-0.002} & -0.002 & 0.002 & -0.05 & 0.947 & \textbf{-0.005} & -0.004 & 0.009 & 0.19 & 0.903 \\ 
  $\beta_2$ & \textbf{0.194} & 0.191 & 0.014 &-0.25& 0.885 & \textbf{0.060} & 0.060 & 0.022 & 0.04&0.887 \\ 
  $\beta_3$ & \textbf{-0.171} & -0.164 & 0.018 & 0.38 & 0.867 & \textbf{0.081} & 0.059 & 0.048 &-0.48 & 0.838 \\ 
  $\sigma_{\beta_1}$ & \textbf{0.010} & 0.010 & 0.003 & -0.04& 0.926 & \textbf{0.008} & 0.012 & 0.017 & 0.22&0.863 \\ 
  $\sigma_{\beta_2}$  & \textbf{0.064} & 0.055 & 0.012 & -0.76&0.634 & \textbf{0.027} & 0.036 & 0.025 & 0.35&0.833 \\ 
  $\sigma_{\beta_3}$  & \textbf{0.079} & 0.077 & 0.014 &-0.13& 0.877 & \textbf{0.068} & 0.077 & 0.027 & 0.33&0.818 \\ 
  $\lambda_1$ & \textbf{362} & 363 & 16.6 &0.08& 0.956 & \textbf{321} & 328 & 60.7 & 0.11&0.918 \\ 
  $\sigma_{\lambda_1}$ & \textbf{93.6} & 97.6 & 12.8 & 0.31&0.931 & \textbf{132} & 142 & 27.9 & 0.36&0.929 \\ 
   $\lambda_2$ & \textbf{643} & 643 & 33.4 & 0.002&0.926 & \textbf{726} & 743 & 69.8 &0.26& 0.904 \\ 
  $\sigma_{\lambda_2}$ & \textbf{149} & 149 & 19.4 & -0.03& 0.970 & \textbf{128} & 144 & 30.0 & 0.53&0.970 \\ 
   \hline
\end{tabular}
\end{table}

A spreadsheet with summary results for each of the $24$ simulation scenarios is available as a supplementary file online.  Additional simulation studies and illustrations are described in the appendices.  In Appendix~\ref{cpsec} a simulation involving up to five changepoints $(K=1,2,3,4,$ or $5$) demonstrates the feasibility  to accurately detect the number and location of a larger number of changepoints. In Appendix~\ref{clussec} a simulation with up to $4$ classes ($C=1,2,3,4$) demonstrates the feasibility to correctly identify the class memberships even when the maximum number of classes is over-specified.  Additional simulation studies illustrate the effect of prior specification on the posterior for key model parameters, which is important to consider for any Bayesian approach.  Appendix~\ref{cpsec} presents a study with different prior choices for the number of changepoints. The correct number of changepoints had the highest posterior probability for all priors considered, but certain prior choices favored over- or under-detection of the changepoints.  Thus, the choice of prior can be used to imply a more conservative model that avoids over-estimating changepoints. In Appendix~\ref{clusalpha}, a study on the effect of the cluster concentration parameter $\alpha$ illustrates how a higher $\alpha$ favors more equality among the latent class proportions in the posterior; however, the accuracy of the class memberships and other model parameters are relatively robust to  changes in $\alpha$. In Appendix~\ref{varpri}, a study using alternative priors for the variances of the random effects reveals that an appropriately scaled half-Cauchy prior performs similarly to the uniform priors used here; generic (unscaled) half-Cauchy priors and uninformative inverse-gamma priors perform poorly and are not recommended.

\section{Implementation}
\label{impl}
We have created a well-documented and user friendly package for R, called {\tt BayesianPGMM}, to estimate the model under the default priors given in Section~\ref{priors}.   This package  provides estimates and 95\% credible intervals for all model parameters, assesses convergence of the algorithm, and includes functions to automatically visualize and summarize results.   The package is available at \url{https://github.com/lockEF/BayesianPGMM}.  A brief tutorial on the use of the package is given in Appendix~\ref{clussec}.  

\section{Discussion}  
\label{discussion}

In this article we have described a mathematical framework for piecewise growth mixture models with unknown number of changepoints, and we have proposed a Bayesian estimation scheme that is appropriate for a wide variety of application scenarios.  The advantages of our proposed method are its \emph{flexibility} and its \emph{simple interpretation}.  The model can accommodate many diverse patterns of growth, with subject-level variation and minimal a priori assumptions on important parameters such as the number and location of the changepoints.  This facilitates interpretation, as the locations at which the functional form of a growth trajectory changes, and the nature of those changes, is invaluable information for many applications in psychometrics and in other fields. 

We have described an application to mouse-tracking data for a language recognition task, which may be extended in several ways.  Here we have focused on modeling the mean mouse trajectories over 16 trials of the same task, across a population of individuals.  Alternatively, one could directly model intra-subject variability over the 16 trials, which would involve a more complex model with multiple hierarchical levels for subjects and trials within a subject.  One could also allow potential covariates of interest, such as mono/bilingual status, to enter the model explicitly in a number of ways.  

By necessity our framework involves many unknown parameters that are influential and highly interdependent,  and this can pose a challenge to their estimation.  In particular, the bias in estimation of class-specific parameters can be non-negligible, especially if there is considerable overlap between the two classes. And the computational time required to fit the model is not trivial for large datasets.  The application described in Section~\ref{app} took approximately $1$ hour to run on a laptop with a 2.5 GHz i7 processor, and computing time scales approximately linearly with the sample size, the number of measurement occasions, and the number of MCMC iterations.  In practice we recommend monitoring convergence closely, and only interpreting parameter estimates after the MCMC algorithm has reached the desired level of convergence.  Fortunately, in certain situations poor convergence may be addressed by increasing the number of MCMC iterations for a given application, if necessary. 

 In the application and simulation we have focused on the model with a maximum of $2$ changepoints ($K=2$), and $2$  classes ($C=2$).  The framework readily allows for more changepoints and classes, and a supplemental document describing additional simulations and illustrations when $K > 2$ or $C > 2$ is available online.  However, we caution against over-parameterization.  At minimum, $K$ must be restricted by the number of time-points $M_i$ that are commonly observed across subjects.  In particular, if the number of active changepoints is greater than $M_i-1$, this requires inferring multiple changepoints between two measurement occasions for subject $i$, which presents identifiability issues.     DIC or other model selection metrics may be used to  select these hyper-parameters. But the asymptotically motivated assumptions for DIC, such as multivariate normality of the posterior, are likely not well satisfied for smaller sample sets.  Diagnostics such as the widely applicable information criterion (WAIC) or  leave-one-out cross-validation (LOO) \citep{vehtari2017practical} may be useful alternatives. WAIC and LOO consider the accuracy of the posterior predictive distribution for held-out data points, and in this context, a ``new data point" may be the entire trajectory for an individual, or a single measurement occasion.  
 
 Our framework may be extended in several ways. In this article we focus on a model with no discontinuity (jumps) in the outcome over time.  We think continuity is a reasonable assumption for most applied situations, but the model may be extended to accommodate discontinuity at the changepoints if necessary.  In this article we have also only considered piecewise segments that have a linear form.  In theory, these segments may take an exponential, quadratic, or other polynomial form.  Allowing these forms to be unknown and vary across different segments would improve the flexibility of the model, but would increase the complexity of estimation and interpretation. 
 
 More advanced and targeted estimation techniques may improve the efficiency and accuracy of posterior computation.  For example, reversible jump MCMC \citep{green1995reversible} inherently infers the presence of fundamentally different parameters,  and may be used in this context to jump between models with different number of changepoints  or different functional forms for each phase during MCMC.  Moreover, there is a considerable literature on Bayesian estimation for changepoint problems involving a single process with multiple phases realized from distinct stationary distributions (see, e.g., \citet{carlin1992hierarchical,chib1998estimation,fearnhead2006exact}).  That is, $y_j \sim p(\eta_j)$ where parameters $\eta_j$ are given by phases over time: 
 \[\eta_j = \begin{cases} 
 \theta_1 \text{ if } x_j < \lambda_1 \\
 \theta_2 \text{ if } \lambda_1 \leq x_j < \lambda_2 \\
 \vdots \\
 \theta_{K+1} \text{ if } \lambda_{K-1} \leq x_j < \lambda_K. 
 \end{cases}\]
In particular, \citet{chib1998estimation}  suggests a hidden Markov approach, that models the probability of transitioning from one phase to the next for discrete time points.  Such an approach is not readily extended to our context, chiefly because the assumption of stationarity within each phase is not satisfied, and modeling segment-wise constant probabilities of change at discrete time points is not readily incorporated into a random effects framework where the location of the changepoint(s) are of substantial interest. However, generally this literature provides many tools that may improve Bayesian estimation of segmented growth models for a population.  Improving the estimation and flexibility of segmented growth models remains an exciting area of active and future research.    

\section*{Acknowledgements} \vspace{-10 pt} We would like to thank Sara Incera and 	Conor T. McLennan of Cleveland State University for graciously providing the mouse-tracking data described in Section~\ref{app}.  This work was supported in part by  NIH grant ULI RR033183/KL2 RR0333182 [to EFL].
   
\vspace{\fill}\pagebreak

\appendix

\section{Multiple change-point simulation}
\label{cpsec}

Here we describe a simulation study in which longitudinal data are generated with anywhere from $0$ to $5$ changepoints, and we assess both accuracy in detecting the number of changepoints and estimation accuracy for the mean location of the changepoints.  All data are simulated according to the following model, which reflects Equation (\ref{eq3}): 
\[y_{ij} = \beta_{i,0}+ \beta_{i,1} x_{ij} + \sum_{k=1}^5 \beta_{i,k+1} (x_{ij}-\lambda_{i,k})^+ \mathbbm{1}_{\{k \leq \mathcal{K}\}}  + \epsilon_{ij},\]  
with
\begin{itemize}
\item $30$ individuals ($i=1,\hdots,30$)
\item $20$ time points ($x_{ij} = 0,\hdots,19$)
\item Intercepts $\beta_{i,0} \sim N(0,0.05)$
\item Potential changepoints $\lambda_{i,1} \sim N(3,\sigma_\lambda^2), \lambda_{i,2} \sim N(6,\sigma_\lambda^2), \lambda_{i,3} \sim N(9,\sigma_\lambda^2), \lambda_{i,4} \sim N(12,\sigma_\lambda^2),\lambda_{i,5} \sim N(15,\sigma_\lambda^2)$
\item Potential slope changes $\beta_{i,k+1} \sim  N\left((-1)^k,0.05\right)$ for $k=0,\hdots,5$
\item Error $\epsilon_{ij} \sim N(0,0.5)$.   
\end{itemize}

The manipulated conditions are the number of changepoints present, $\mathcal{K} = \{0,1,2,3,4,5\}$, and the standard deviation of the individual changepoints, $\sigma_\lambda = \{0.2,0.5\}$.  

We generate $30$ replicated datasets for each combination of the initial conditions, yielding $30 \times 6 \times 2=360$ simulated datasets.  For each simulation, we estimate the model as described in Section~\ref{inference} with a maximum of $5$ possible changepoints ($K=5$).  The prior for the number of changepoints is specified as $\mathcal{K} \sim \text{Binomial}(5,0.5)$, where $5$ is the number of potential changepoints and $0.5$ is the probability of including each changepoint.  

\begin{table}[!htb]
\caption{True number of changepoints ($\mathcal{K}$) and mean posterior probability $\hat{\mathcal{K}}$.}
\label{suptab1}
\centering
\begin{tabular}{|r|rrrrrr|}
\hline
$\mathbf{\sigma_\lambda = 0.2}$& $\mathcal{K} =0$ & $\mathcal{K}=1$ & $\mathcal{K}=2$ & $\mathcal{K}=3$ & $\mathcal{K}=4$ & $\mathcal{K}=5$\\	
\hline
$\hat{\mathcal{K}}=0$ & \textbf{1.000} & 0.00& 0.000 & 0.000 & 0.000 & 0.000 \\
$\hat{\mathcal{K}}=1$ &0.000 & \textbf{0.977} & 0.000 & 0.000 & 0.000 & 0.000\\
$\hat{\mathcal{K}}=2$ & 0.000 & 0.022 & \textbf{0.996} & 0.000 & 0.000 & 0.000 \\
$\hat{\mathcal{K}}=3$ & 0.000 & 0.000& 0.004 & \textbf{0.987} & 0.000& 0.000\\
$\hat{\mathcal{K}}=4$ &0.000 & 0.000& 0.000 & 0.0134 & \textbf{0.999} & 0.011\\
$\hat{\mathcal{K}}=5$ & 0.000 & 0.000 & 0.00 & 0.000 & 0.001 & \textbf{0.989}\\
\hline
\hline
$\mathbf{\sigma_\lambda = 0.5}$ & $\mathcal{K} =0$ & $\mathcal{K}=1$ & $\mathcal{K}=2$ & $\mathcal{K}=3$ & $\mathcal{K}=4$ & $\mathcal{K}=5$\\	
\hline
$\hat{\mathcal{K}}=0$ & \textbf{1.000} & 0.00& 0.000 & 0.000 & 0.000 & 0.000 \\
$\hat{\mathcal{K}}=1$ &0.000 & \textbf{0.933} & 0.000 & 0.000 & 0.000 & 0.000\\
$\hat{\mathcal{K}}=2$ & 0.000 & 0.067 & \textbf{0.998} & 0.000 & 0.054 & 0.022 \\
$\hat{\mathcal{K}}=3$ & 0.000 & 0.000& 0.002 & \textbf{0.997} & 0.080 & 0.343\\
$\hat{\mathcal{K}}=4$ &0.000 & 0.000& 0.000 & 0.003 & \textbf{0.867} & 0.113\\
$\hat{\mathcal{K}}=5$ & 0.000 & 0.000 & 0.00 & 0.000 & 0.000 & \textbf{0.522}\\
\hline
\end{tabular}
\end{table}

The resulting posterior distributions for $\mathcal{K}$ are shown in Table~\ref{suptab1}, averaged over the $30$ replications for each cell.  When $\sigma_\lambda=0.2$ the correct number of changepoints, is generally recovered correctly with high posterior probability; the true number of changepoints always has average posterior probability $>0.95$.  When $\sigma_\lambda=0.5$ the true number of changepoints has the highest average posterior probability for all cases, but is frequently underestimated when there are $4$ or $5$ changepoints.  

\begin{figure}[!htb]
\begin{center}
\includegraphics[width = \textwidth]{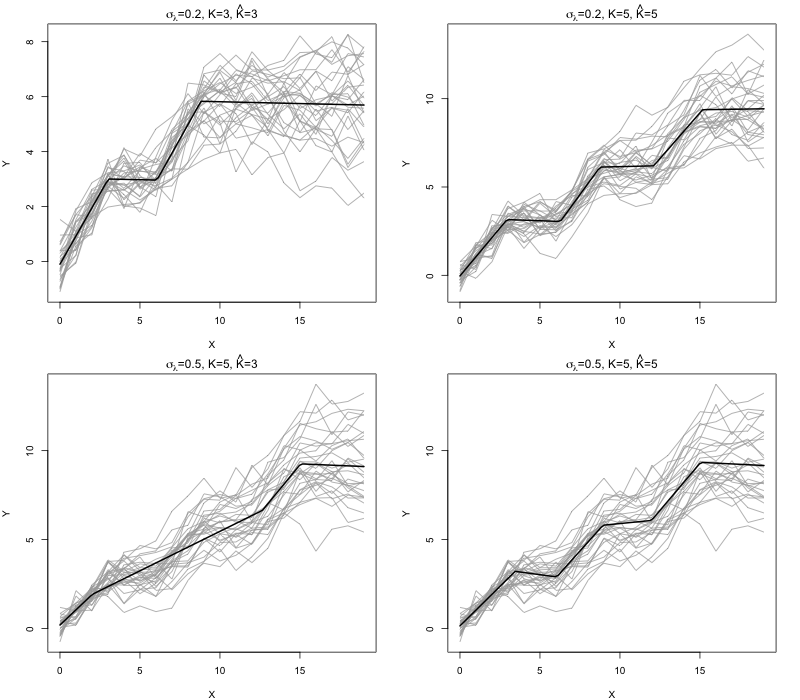}
\caption{Spaghetti plot of individual trajectories (gray) and posterior estimates (black) for select simulations.} 
\label{SpagPlotCPs}
\end{center}
\end{figure}

The lack of precision in detecting several changepoints when $\sigma_\lambda=0.5$, evident in Table~\ref{tab1}, is because the higher variance of the changepoints makes them more difficult to distinguish.   This is apparent in Figure~\ref{SpagPlotCPs}, which illustrates the data and resulting mean model fit for different situations.  The bottom two panels show two different models for the same dataset with $5$ changepoints and $\sigma_\lambda=0.5$.  The bottom left panel shows the posterior mean when the number of changepoints is under specified as $\hat{\mathcal{K}}=3$; this under specified model has substantial posterior probability, illustrating the difficulty in detecting the true number of changepoints.  The  posterior mean for the same dataset under the correct number of changepoints ($\hat{\mathcal{K}}=5$) is shown in the bottom right panel. 

Table~\ref{tabCP} shows the estimated mean changepoint locations, when the number of changepoints are correctly specified, for each of the $10$ simulation scenarios that involve at least one changepoint ($\mathcal{K}=1,2,3,4,5$; $\sigma_\lambda = 0.2,0.5$).  For each scenario the recovery of the changepoints are generally accurate.  The accuracy in estimating each changepoint does not suffer greatly as the number of changepoints increase.  

\begin{table}[!htb]
\caption{Mean changepoint locations (and standard deviation of replications) for the $\mathcal{K}=1, 2, 3, 4,$ and $5$ changepoint scenarios, for posterior draws where the number of changepoints are correctly detected.}
\label{tabCP}
\centering
\begin{tabular}{|r|ccccc|}
\hline
$\mathbf{\sigma_\lambda = 0.2}$ & $\mathcal{K}=1$ & $\mathcal{K}=2$ & $\mathcal{K}=3$ & $\mathcal{K}=4$ & $\mathcal{K}=5$\\	
\hline
$\lambda_1=3$ & 2.96 (0.14) & 2.99 (0.16) & 3.01 (0.14) & 3.03 (0.18) & 2.98 (0.18) \\
$\lambda_2=6$ & & 5.94 (0.12)  & 5.97 (0.17) & 6.01 (0.16) & 5.99 (0.17) \\
$\lambda_3=9$ & &  & 8.98 (0.15) & 8.99 (0.17) & 8.99 (0.20) \\
$\lambda_4=12$ & & &    & 11.99 (0.18) & 12.00 (0.20)\\
$\lambda_5=15$ & & &   & & 15.00 (0.16)\\
\hline 
\hline
$\mathbf{\sigma_\lambda = 0.5}$ & $\mathcal{K}=1$ & $\mathcal{K}=2$ & $\mathcal{K}=3$ & $\mathcal{K}=4$ & $\mathcal{K}=5$\\	
\hline
$\lambda_1=3$ & 3.01 (0.11) & 2.99 (0.20) & 2.97 (0.20) & 3.02 (0.18) & 2.92 (0.22) \\
$\lambda_2=6$ & & 6.03 (0.16)  & 5.92 (0.21) & 6.02 (0.23) & 5.99 (0.21) \\
$\lambda_3=9$ & &  & 9.07 (0.21) & 8.95 (0.21) & 8.98 (0.22) \\
$\lambda_4=12$ & & &    & 12.03 (0.18) & 12.03 (0.21)\\
$\lambda_5=15$ & & &   & & 14.96 (0.27)\\
\hline
\end{tabular}
\end{table}

To assess the effect of prior specification on the posterior number of changepoints, we repeat the entire simulation above with alternative priors $\mathcal{K} \sim \text{Binomial}(5,0.25)$ or $\mathcal{K} \sim \text{Binomial}(5,0.75)$.  The results are summarized in Table~\ref{tabCPprior}, which shows the average posterior probability across simulation scenarios under-estimating the true number of changepoints $\mathcal{K}$ ($\hat{\mathcal{K}}<\mathcal{K}$), correctly estimating $\mathcal{K}$ ($\hat{\mathcal{K}}=\mathcal{K}$), and over-estimating $\mathcal{K}$ ($\hat{\mathcal{K}}>\mathcal{K}$), for the different prior specifications.  The correct number of changepoints has the highest posterior probability in all scenarios, but as expected smaller values of the prior binomial probability $p$ tend to bias the results toward under-estimation, and larger values bias the result toward over-estimation.  Thus, for a more conservative prior that will avoid over-detecting  changepoints, one can use a binomial prior with a small probability hyper-parameter.    

\begin{table}[!htb]
\caption{Average posterior probability across simulation scenarios under-estimating the true number of changepoints $\mathcal{K}$ ($\hat{\mathcal{K}}<\mathcal{K}$), correctly estimating $\mathcal{K}$ ($\hat{\mathcal{K}}=\mathcal{K}$), and over-estimating $\mathcal{K}$ ($\hat{\mathcal{K}}>\mathcal{K}$), for different prior specifications.  }
\label{tabCPprior}
\centering
\begin{tabular}{|c|ccc|}
\hline
$\mathbf{\sigma_\lambda = 0.2}$& Under-estimation  & Correct estimation & Over-estimation \\	
\hline
$\mathcal{K} \sim \mbox{Binom}(5,\mathbf{0.25})$ & 0.367 & 0.633& 0.001 \\
$\mathcal{K} \sim \mbox{Binom}(5,\mathbf{0.5})$ &0.002 & 0.991 & 0.007  \\
$\mathcal{K} \sim \mbox{Binom}(5,\mathbf{0.75})$ & 0.006 & 0.883 & 0.111  \\
\hline
\hline
$\mathbf{\sigma_\lambda = 0.5}$ & Under-estimation  & Correct estimation & Over-estimation\\	
\hline
$\mathcal{K} \sim \mbox{Binom}(5,\mathbf{0.25})$ & 0.441 & 0.559 & 0.000  \\
$\mathcal{K} \sim \mbox{Binom}(5,\mathbf{0.5})$ &0.10 & 0.886 & 0.012 \\
$\mathcal{K} \sim \mbox{Binom}(5,\mathbf{0.75})$ & 0.027 & 0.877 & 0.095  \\
\hline
\end{tabular}
\end{table}

\section{Multi-class illustration}
\label{clussec}

Here we describe a simple example to illustrate the clustering properties and simultaneous changepoint detection of the {\tt BayesianPGMM} package.  This section also serves as  a brief tutorial, with commands to reproduce the results below after the package is installed and loaded to the R workspace. 

\begin{figure}[ht]
\begin{center}
\includegraphics[width = \textwidth]{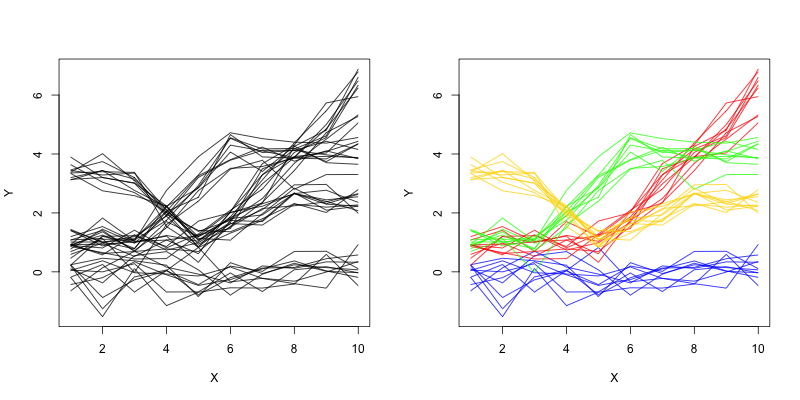}
\caption{Spaghetti plot of generated data without showing classes (left) and colored by latent classes (right). The blue class has $0$ changepoints, the red has $1$ changepoint, the green has $2$ changepoints, and the gold has $3$ changepoints. } 
\label{SpagPlotClusts}
\end{center}
\end{figure}

We generate the data shown in Figure~\ref{SpagPlotClusts}. These data can be loaded in R via the command 
{\tt data(SimData4classes)} after the package is installed, and can be visualized as shown using the command {\tt plotPGMM(X,Y)}.   These data consist of four latent classes, each with ten individuals with measurements for the same $10$ time points.  Each latent class has a different number of changepoints $0$, $1$, $2$, or $3$.

\begin{figure}[!h]
\begin{center}
\includegraphics[width = \textwidth]{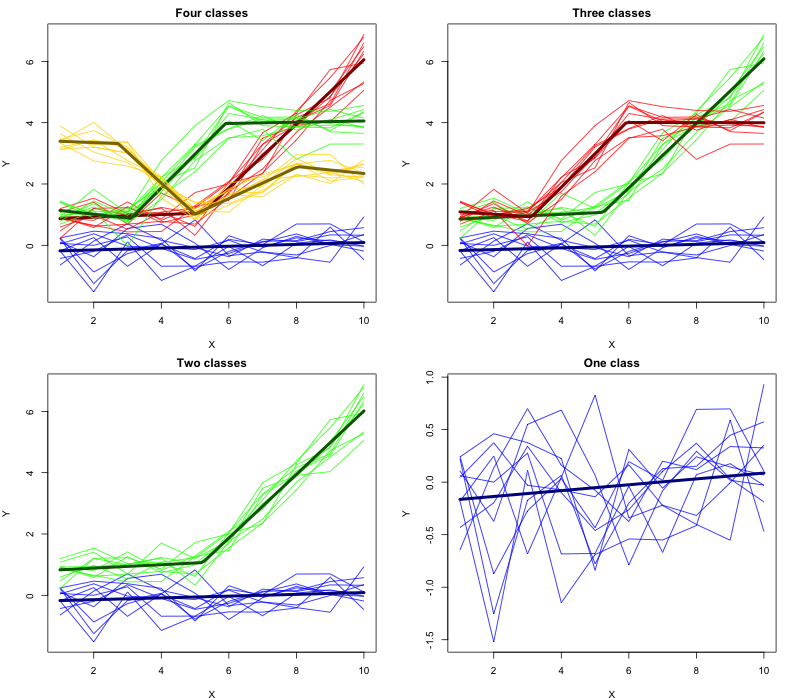}
\caption{Spaghetti plot of the simulated data with different number of latent classes present, with colors showing the estimated class clustering. The trajectory defined by the mean parameters for each class are shown in \textbf{bold}. } 
\label{SpagPlotClustsColor}
\end{center}
\end{figure}

We estimate the posterior model with four latent classes, and up to $3$ changepoints in each class, using the command 
\begin{verbatim} Fit <- BayesPGMM(X,Y,max_cp=3,n_clust=4) .\end{verbatim}  The resulting class clustering and mean fits can be visualized using the command {\tt plotPGMM(X,Y,Fit)}, as shown in the top left panel of Figure~\ref{SpagPlotClustsColor}.  The resulting clustering matches the true latent classes, and the correct number of changepoints are detected for each class.

To illustrate robustness to over-specification of the number of classes we also fit the model with the same specification (four classes, $3$ potential changepoints) to a reduced dataset with one latent class removed.  Specifically, we remove the $10$ individuals belonging to the fourth class, leaving three latent classes with $0$, $1$ and $2$ changepoints.  Thus, we use a four class model to estimate data with three classes:  \begin{verbatim} Fit <- BayesPGMM(X[1:30,],Y[1:30,],max_cp=3,n_clust=4) .\end{verbatim} The results can again be visualized using {\tt plotPGMM(X[1:30,],Y[1:30,],Fit)}, as in the top right panel of Figure~\ref{SpagPlotClustsColor}.  The latent classes and number of changepoints in each class are again recovered correctly. In particular, only three of the possible four latent classes are represented, leaving the extraneous fourth class empty.  We similarly fit the model with four classes and $3$ potential changepoints to data with only two of the classes (with $0$ and $1$ changepoints), and to data with only one class (with $0$ changepoints).  The results, shown in the bottom two panels of Figure~\ref{SpagPlotClustsColor}, again recover the true clustering and number of changepoints, leaving extraneous classes empty.

For each of the four simulated datasets above, the recovery of the clustering and true number of changepoints were validated with $10$ independent replications.  

\section{Clustering prior simulation}
\label{clusalpha}

Here we describe a simulation to illustrate the effect of the concentration parameters for the Dirichlet clustering prior on the posterior.  It is common to set each value of the $C$-dimensional concentration parameter, where $C$ is the number of clusters, to a constant $\alpha$: Dirichlet($\alpha$, ..., $\alpha$).  Smaller values of $\alpha$ suggest less parity in the class sizes (e.g., one class is much larger than the other), while larger values of $\alpha$ suggest more parity in the class sizes.  To illustrate, we consider a two-class model, for which the Dirichlet($\alpha,\alpha$) distribution is equivalent to a Beta$(\alpha,\alpha)$  distribution for the proportion of one class.  By default we use $\alpha=1$, which is equivalent to a Uniform$(0,1)$ distribution; more generally, a Dirichlet$(1,\hdots,1)$ distribution is uniform over the unit simplex.

\begin{figure}[!h]
\begin{center}
\includegraphics[width = 0.75\textwidth]{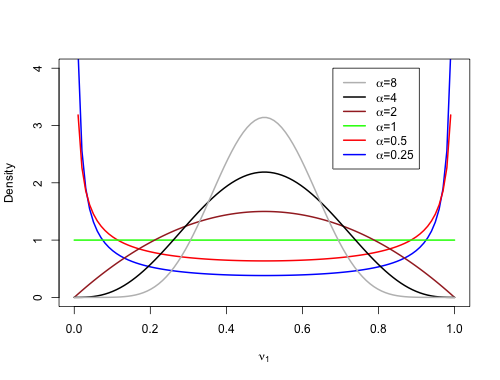}
\caption{Prior density of $\nu_1$ for different Dirichlet($\alpha,\alpha$) distributions. } 
\label{BetaPlots}
\end{center}
\end{figure}

Herein in addition to the two-class model with $\alpha=1$, we consider Dirichlet priors with $\alpha=0.25$, $\alpha=0.5$, $\alpha=1$, $\alpha=2$, $\alpha=4$, and $\alpha=8$.  The resulting prior distributions for a single class probability $\nu_1$ ($\nu_2=1-\nu_1$) are shown in Figure~\ref{BetaPlots}.  Note that $\alpha=0.5$ corresponds to a Jeffrey's prior \citep{jeffreys1946invariant}.  

We simulate an additional $100$ realizations of the simulation scheme in Section~\ref{sims}, under the application-motivated scenario with $N=60$, $M_i=50$, $\nu_1=0.80$, and $\mathcal{K}_2=2$.  We compute the posterior for each of $\alpha=\{0.25,0.5,2,4,8\}$ for $20$ realizations, with otherwise the same settings as those used in the main simulation with $\alpha=1$.  The average of the posterior means for the latent proportion of the smaller class $\nu_2=0.2$ is shown for each value of $\alpha$ in Table~\ref{tab1}.  The estimated latent proportion tends to increase above $0.2$ for higher values of $\alpha$;  this is expected, as higher values of $\alpha$ tends to bias estimates toward equal class proportions.  However, this appears to have little affect on the overall accuracy of the posterior: the misallocation rate of the latent class memberships is not substantially affected (Table~\ref{tabDir}), and the posterior accuracy of other model parameters are also not substantially affected (Table~\ref{suptab2}).   

\begin{table}[!htb]
\caption{Mean class $2$ proportion $\nu_2$ $(\nu_2=0.2)$, and mean class misallocation rate, for Dirichlet concentration parameter $\alpha$. }
\label{tabDir}
\centering
\begin{tabular}{r|ccccccc}

  & $\alpha=0.25$ & $\alpha=0.5$ & $\alpha=1$ & $\alpha=2$ & $\alpha=4$ & $\alpha=8$\\ 
 \hline
 \textbf{$\hat{\nu}_2$} &  0.18 & 0.19 & 0.21 & 0.26 & 0.26 & 0.31 \\
 Misallocation & 0.15 & 0.13 & 0.12 & 0.10 & 0.12 & 0.12\\
  \hline
\end{tabular}
\end{table}  
 
\begin{table}[!htb]
\caption{Summary of mean parameter estimates in Class 1 and 2 for concentration parameter $\alpha<1$ ($\alpha=0.25$ or $\alpha=0.5$), $\alpha=1$, or $\alpha>1$ ($\alpha=2$ or $\alpha=4$ or $\alpha=8$). }
\label{suptab2}
\centering
\begin{tabular}{rrrrrrrrr}
  \hline
& \textbf{Class 1} & $\alpha<1$ & $\alpha=1$ & $\alpha>1$ & \textbf{Class 2} & $\alpha<1$ & $\alpha=1$ & $\alpha>1$\\ 
  \hline
$\sigma_\epsilon$ & \textbf{3.16} & 3.17 & 3.17 & 3.18 & \textbf{3.16} & 3.17 & 3.17 & 3.18  \\ 
  $\beta_1$ & \textbf{-0.002} & -0.003 & -0.003 & -0.003 & \textbf{-0.005} & -0.0009 &	-0.0008	& -0.0008 \\ 
  $\beta_2$ & \textbf{0.194} & 0.185 & 0.186 & 0.192& \textbf{0.060} & 0.068 	&0.066&	0.062   \\ 
  $\beta_3$ & \textbf{-0.171} & -0.153	&-0.154	&-0.164 &\textbf{0.081} & 0.0217	& 0.0325& 	0.0434  \\ 
  $\sigma_{\beta_1}$ & \textbf{0.010} & 0.010	& 0.010 &	0.010 & \textbf{0.008} & 0.033	&0.025 &	0.016  \\ 
  $\sigma_{\beta_2}$  & \textbf{0.064} & 0.058 	&0.057 &	0.053 & \textbf{0.027} & 0.062	& 0.059& 	0.054  \\ 
  $\sigma_{\beta_3}$  & \textbf{0.079} & 0.089 & 0.087	 &0.081 &  \textbf{0.068} & 0.105 & 	0.099 &	0.090 \\ 
  $\lambda_1$ & \textbf{362} & 362 &	363	& 363 & \textbf{321} & 328 & 	327	& 328 \\ 
  $\sigma_{\lambda_1}$ & \textbf{93.6} & 98.7	&98.2 & 	95.1 & \textbf{132} & 136	& 143 &151 \\ 
   $\lambda_2$ & \textbf{643} & 650 &	650	& 645 & \textbf{726} & 719 & 708	& 708\\ 
  $\sigma_{\lambda_2}$ & \textbf{149} & 147 &	146	& 144  & \textbf{128} & 140 & 147	 & 161\\ 
   \hline
\end{tabular}

\end{table}

\section{Variance prior simulation}
\label{varpri}

Here we describe a simulation in which we consider alternative priors for the variance (or standard deviation) of the random effects.  By default we have used a uniform prior for the standard deviation, with a lower bound of $0$ and an upper bound that depends on the context of the parameter (see Section~\ref{priors}).  An alternative prior for the standard deviation is the half-Cauchy prior \citep{polson2012half}, which is a Cauchy distribution truncated above $0$: 
\[p(x \mid \gamma) = \frac{2}{\pi \gamma \left( 1+(x/\gamma)^2\right)} \; \; \text{for } x>0,\]   
where $\gamma$ is a scale parameter.  We implement the half-Cauchy prior for all random-effects parameters $\left(\{\sigma_{c,\beta_k}^2\},\{\sigma_{c,\lambda_k}^2\}\right)$, and under two different strategies to select $\gamma$, 
\begin{enumerate}
\item \textbf{Scaled}, in which $\gamma$ depends on the parameter.  Here, $\gamma$ is selected such that the $90$th percentile of the resulting half-Cauchy distribution is given by the upper bound used for the default uniform distribution.  For example, under the default uniform prior  $\sigma_{c,\lambda_1} \overset{iid}{\sim} \mbox{Uniform}(0,b)$  where $b = \frac{\mbox{max}(X)-\mbox{min}(X)}{4}$, while under the scaled Cauchy prior $P(\sigma_{c,\lambda_1} < b)=0.9$.
\item \textbf{Unscaled}, in which $\gamma=25$ for all parameters; this is suggested as the default half-Cauchy prior for a scale parameter in the {\tt laplacesDemon} R package \citep{statisticat2015laplacesdemon}.  
\end{enumerate}
As another alternative, we consider an $IG(0.001,0.001)$ distribution for the variances of the random effects.  

We repeat $100$ simulations from Section~\ref{sims}, under the application-motivated scenario with $N=60$, $M_i=50$, $\nu_1=0.80$, and $\mathcal{K}_2=2$.  For each replication we consider, in addition to the default uniform prior, a the scaled half-Cauchy prior, unscaled half-Cauchy prior, and inverse-gamma prior for the random effects.  

The diffuse inverse-gamma prior (here IG$(0.001,0.001)$) is generally not recommended for modeling the variance of hierarchical random effects \citep{gelman2006prior}, partly because its density is unstable as the variance approaches $0$.  This is especially worrisome when the number of random effects are small, or in the context of mixture models, where the number of observations within a class may be small and vary during posterior sampling.  Indeed, our implementation of IG$(0.001,0.001)$ priors failed during posterior sampling for each replication, because of numerical errors caused by extreme values.  

The resulting average parameter estimates for the uniform, scaled half-Cauchy, and unscaled half-Cauchy priors are shown in Table~\ref{suptab3}.  The results for the scaled half-Cauchy priors are mostly comparable to the results for the default uniform priors, although mean estimates for the standard deviations for the changepoint locations in Class 2 (the smaller class) are inflated.  For the unscaled half-Cauchy priors the posterior standard deviations for the random coefficients in Class 2 are highly inflated, and other parameter estimates for Class 2 are generally less accurate.  These results demonstrate that appropriate scaling of the prior for hierarchical random effects is important, especially for the accurate identification of latent classes that have a small number of individuals. 

\begin{table}[!htb]
\caption{Summary of mean parameter estimates in Class 1 and 2 with different prior choices for the model random effects, including the default uniform priors, scaled half-Cauchy and unscaled half-Cauchy (HC$(0,25)$). }
\label{suptab3}
\centering
\begin{tabular}{rrrrrrrrr}
  \hline
& \textbf{Class 1} & Uniform & Scaled HC & HC$(0,25)$& \textbf{Class 2} & Uniform & Scaled HC & HC$(0,25)$\\ 
  \hline
$\sigma_\epsilon$ & \textbf{3.16} & 3.17 & 3.18 & 3.19 & \textbf{3.16}  & 3.17 & 3.18& 3.19  \\ 
  $\beta_1$ & \textbf{-0.002}  & -0.003 & -0.003 & -0.003 & \textbf{-0.005}  &	-0.0008	& -0.002& -0.001\\ 
  $\beta_2$ & \textbf{0.194}  & 0.186 & 0.190& 0.178 & \textbf{0.060}  	&0.066&	0.070& 0.045 \\ 
  $\beta_3$ & \textbf{-0.171} 	&-0.154	& -0.159& -0.142 &\textbf{0.081} 	& 0.0325& 0.034	&0.028\\ 
  $\sigma_{\beta_1}$ & \textbf{0.010} 	& 0.010 &0.010	&0.010 & \textbf{0.008} 	&0.025 &0.017	& 5.12\\ 
  $\sigma_{\beta_2}$  & \textbf{0.064} 	&0.057 &	0.054 & 0.065& \textbf{0.027} 	& 0.059& 0.027	& 8.61  \\ 
  $\sigma_{\beta_3}$  & \textbf{0.079}  & 0.087	 & 0.082& 0.103&  \textbf{0.068}  & 	0.099 &	0.081& 8.51 \\ 
  $\lambda_1$ & \textbf{362}  &	363	& 362& 362& \textbf{321}  & 	327	& 335&337 \\ 
  $\sigma_{\lambda_1}$ & \textbf{93.6} 	&98.2 & 95.1	& 97.1& \textbf{132} 	& 143 & 263&176 \\ 
   $\lambda_2$ & \textbf{643}  &	650	& 646& 654& \textbf{726}  & 708	& 711& 710\\ 
  $\sigma_{\lambda_2}$ & \textbf{149}  &	146	& 142&  142& \textbf{128}  & 147	 & 235&145\\ 
   \hline
\end{tabular}
\end{table}   

In the {\tt BayesianPGMM} package, we have implemented the scaled half-Cauchy prior as an option, in addition to the default uniform prior. The half-Cauchy has the advantage of not having a hard constraint (e.g., as in the uniform upper bound) and facilitating some shrinkage; however, the half-Cauchy can give non-trivial probability to unreasonably large standard deviations in the right tail of the distribution, and the uniform prior has the advantage of being simple to interpret.   


\newpage
\bibliographystyle{apa}
\bibliography{Piecewise.bib}



%




\end{document}